\documentstyle[twoside,fleqn,espcrc2,epsf,rotate]{article}

\title{$SU(3)$ lattice gauge autocorrelations with anisotropic action}

\author{Terrence Draper, 
        Constantine Nenkov\thanks{Presented by C.\ Nenkov at
         Lattice '96, St.\ Louis.},  
    and Mike Peardon
        \address{Department of Physics and Astronomy,
                 University of Kentucky,
                 Lexington, KY 40506, USA}
        \thanks{This work is supported in part by the U.S. Department
                of Energy under grant numbers DE-FG05-84ER40154 and 
                DE-FC02-91ER75661, and by the Center for
                Computational Sciences, University of Kentucky.}}

\begin{document}

\begin{abstract}

We report results of autocorrelation measurements in pure $SU(3)$ lattice gauge
theory. The computations are performed on the {\sc convex spp1200} parallel
platform within the {\sc canopy} programming environment. The focus of our
analysis is on typical autocorrelation times and optimization of the mixing
ratio between overrelaxation and pseudo-heatbath sweeps for generating gauge
field configurations. We study second order tadpole-improved approximations of
the Wilson action in the gluon sector, which offers the advantage of working on
smaller lattices ($8^3~\times~16$ and $6^3~\times$~12~--~30).  We also make use
of anisotropic lattices, with temporal lattice spacing smaller than the spatial
spacing, which prove useful for calculating noisy correlation functions with
large spatial lattice discretization (of the order of 0.4 fm).

\end{abstract}

\maketitle

\section{NOTATION}

\begin{itemize}
 \item s -- spatial
 \item t -- temporal 
 \item $\tau_{\rm int}$ -- integrated autocorrelation time
 \item $\xi$ -- correlation length
 \item $a_{s}$ -- spatial lattice spacing
 \item $a_{t}$ -- temporal lattice spacing
 \item $\rho = a_{t}/a_{s}$ -- anisotropy ratio
 \item $u_{s}$ -- tadpole-improved mean spatial link
 \item $u_{t}$ -- tadpole-improved mean temporal link
 \item $S_{nd}$ -- lattice action without doublers
 \item Pss -- spatial plaquette 
 \item Pst -- temporal plaquette
 \item Rss -- spatial $2 \times 1$ rectangle 
 \item Rst1 -- temporal $2 \times 1$ rectangle of type-I
 \item Rst2 -- temporal $2 \times 1$ rectangle of type-II
 \item Act -- average gluon action
 \item H-B -- heatbath
 \item Over. -- overrelaxation
 \item MC -- Monte Carlo
\end{itemize}

\section{MOTIVATION}

Recently it has become apparent that coarse anisotropic lattices with $\rho <
1$ can be very useful in performing accurate Monte Carlo simulations of QCD at
low computational cost. This is especially true when modeling heavy quantum
states of QCD -- like glueballs, for example.  Because of the exponential fall
off of the signal, small $a_{t}$ gives better resolution of the correlators at
an early time step. On the other hand, $a_{s}$ should be kept relatively large
because of critical slowing down.
 
Since the anisotropy of the lattice breaks the Euclidean invariance of the
continuum theory, it induces temporal $\xi_{t}^{\rm lat}$ and spatial
$\xi_{s}^{\rm lat}$ correlation lengths which scale as
\begin{equation}
  \frac{\xi_{t}^{\rm lat}}{\xi_{s}^{\rm lat}} =
  \frac{a_{s}}{a_{t}} = \rho^{-1}
\end{equation}
so that $a_{t} \ll a_{s} \Rightarrow \xi_{t}^{\rm lat} \gg \xi_{s}^{\rm lat}$.

On the other hand, the autocorrelations in Monte Carlo updates are proportional
to a power of the correlation length
\begin{equation}\label{pow}
  \tau_{\rm op}\ \propto\ \left( \xi^{\rm lat}_{\rm relevent\ op.} \right)^n,
\end{equation}
where theoretically $n=2$ for local stochastic updates and $n=0$ for
cluster/overrelaxation updates.

In practice, different lattice operators will have very different
autocorrelation times, and we expect operators that couple strongly to
$\xi_{t}^{\rm lat}$ to have larger $\tau_{\rm int}$. Unfortunately
``interesting operators'' in lattice QCD, like those for glueballs, live in the
spatial domain and scale with $\xi^{\rm lat}_{t}$ which is large.

This work tries to address the issue of the scaling behavior of different gluon
operators on anisotropic lattices, and their relevance to the problem of MC
algorithm optimization.

\section{THE GLUON ACTION}

The doubler-free gluon action used in this study is given by
\begin{eqnarray}
S_{nd} & = & -\beta \sum_{x,s>s'}\left(\frac{a_{t}}{a_{s}}\right)
              \left(\frac{5}{3}\frac{P_{ss'}}{u^{4}_{s}}\right) \nonumber \\
       &   & +\beta \sum_{x,s>s'}\left(\frac{a_{t}}{a_{s}}\right)
              \left( \frac{1}{12}\frac{R_{ss'}}{u^{6}_{s}}
             +\frac{1}{12}\frac{R_{s's}}{u^{6}_{s}} \right) \nonumber \\
       &   & -\beta \sum_{x,s}\left(\frac{a_{s}}{a_{t}}\right)
              \left( \frac{4}{3}\frac{P_{st}}{u^{2}_{s}u^{2}_{t}}
             -\frac{1}{12}\frac{R_{st}}{u^{4}_{s}u^{2}_{t}} \right)
\end{eqnarray}
This is a Symanzik-improved lattice action~\cite{sym}, which is accurate up to
errors $O(a_s^4,a_t^2)$ (classically).

The tadpole-improvement scheme is
\begin{eqnarray}
  u_{t} & = & 1 \nonumber \\
  u_{s} & = & \langle P_{ss'} \rangle^{1/4}
\end{eqnarray}
because of the assumption $a_{t} \ll a_{s}$~\cite{al1,al2}.

We examine anisotropic lattices with spatial lattice spacings $a_{s}$ from 0.25
to 0.4 fm and anisotropy $\rho$ ranging from 1/2 to 1/5. The mean spatial
tadpole-improvement factor $u_{s}$ varies from 0.772 to 0.807. For every
lattice considered we generate a corresponding Markov chain of 4000
configurations ({\it i.e.\/} MC sweeps), which are viewed as Monte Carlo time
series of appropriate lattice operators.

\begin{figure}[htb]
  \begin{center}
    \epsfysize=\hsize 
    \leavevmode
    \rotate[l]{\epsfbox{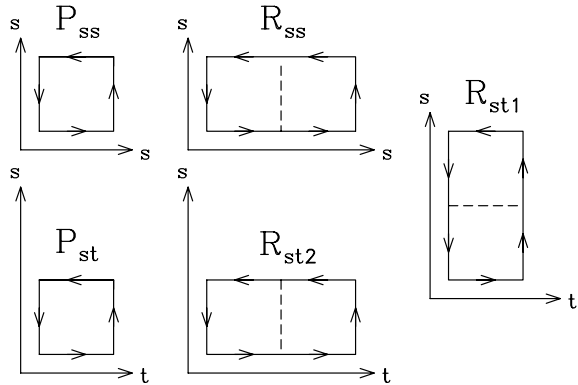}}
  \end{center}
  \caption{Types of planar loops used in this study.}
  \label{plqs}
\end{figure}

On these time series we perform autocorrelation measurements on six lattice
observables: Pss, Pst, Rss, Rst1, Rst2 and Act, which are shown in
Fig.~\ref{plqs}.

%

\section{AUTOCORRELATION FUNCTIONS}

\begin{figure}[htb]
  \begin{minipage}[t]{\hsize}
    \begin{center}
      \leavevmode
      \epsfysize=\hsize
      \rotate[r]{\epsfbox{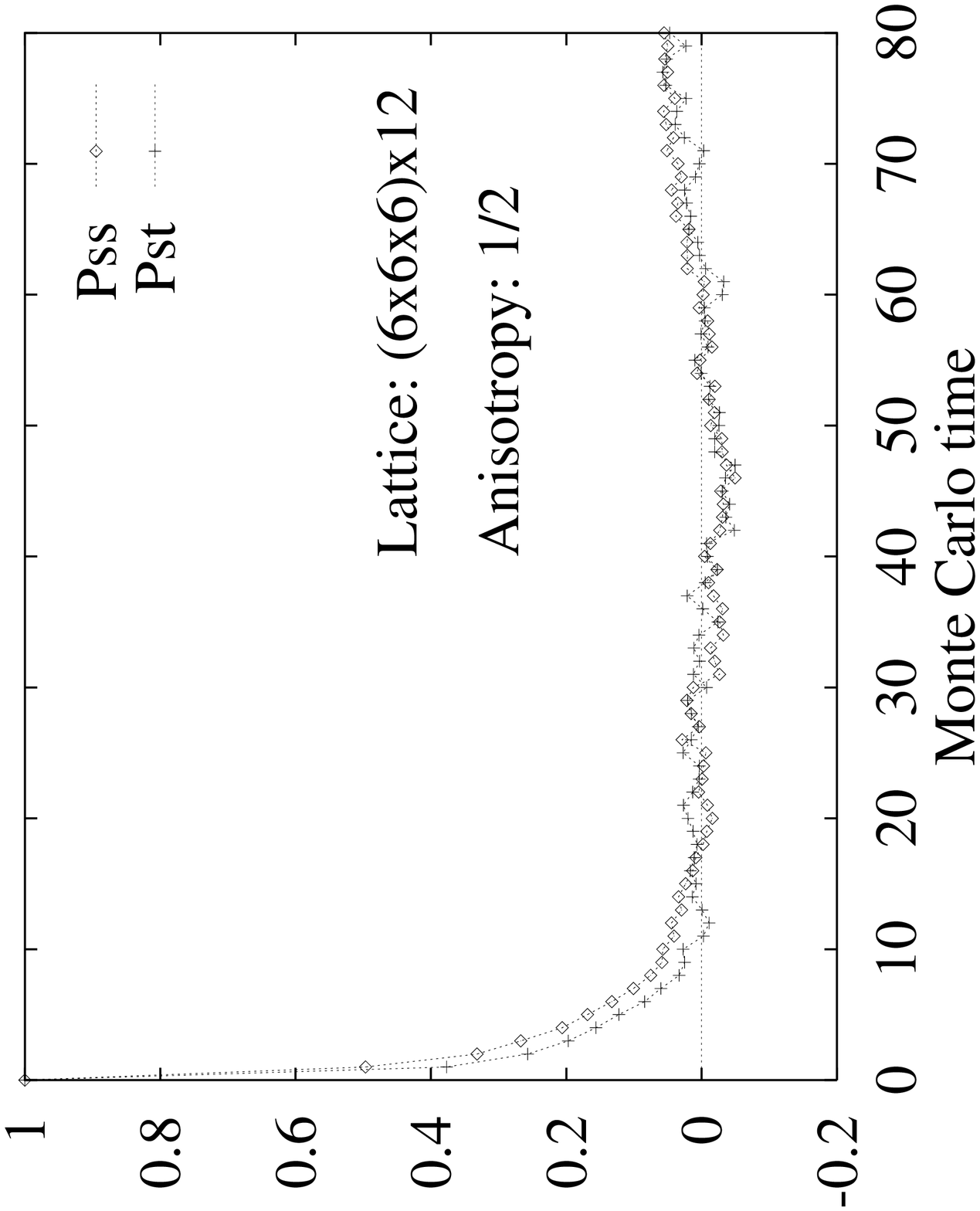}}
    \end{center}
  \end{minipage} 
  \begin{minipage}[t]{\hsize}
    \begin{center}
      \leavevmode
      \epsfysize=\hsize 
      \rotate[r]{\epsfbox{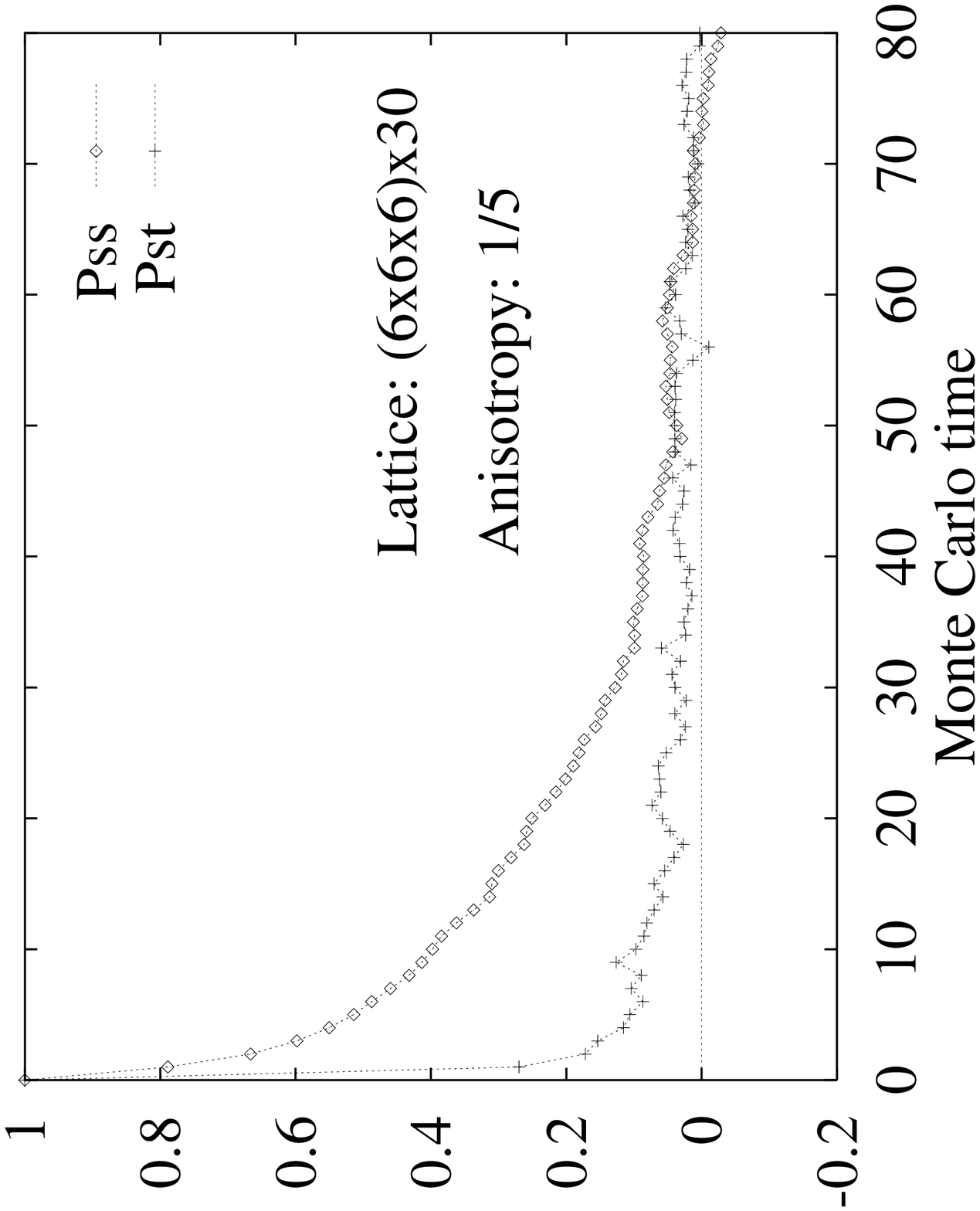}}
    \end{center}
  \end{minipage}
  \caption{Effects of anisotropy on the autocorrelation functions for the 
           plaquettes.}
  \label{awil}
\end{figure}

\begin{figure}[htb]
  \begin{minipage}[t]{\hsize}
    \begin{center}
      \leavevmode
      \epsfysize=\hsize 
      \rotate[r]{\epsfbox{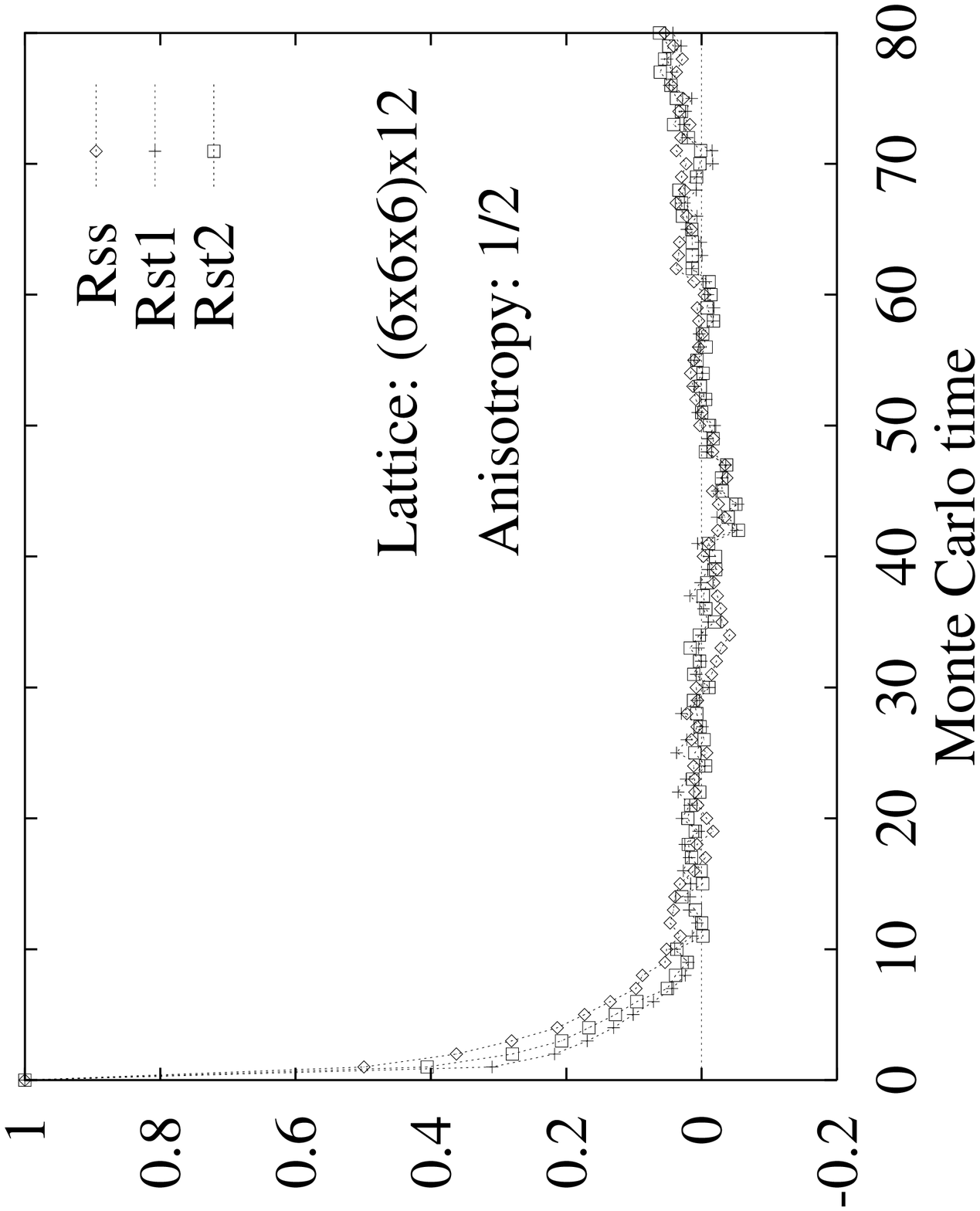}}
    \end{center}
  \end{minipage}
  \begin{minipage}[t]{\hsize}
    \begin{center}
      \leavevmode
      \epsfysize=\hsize 
      \rotate[r]{\epsfbox{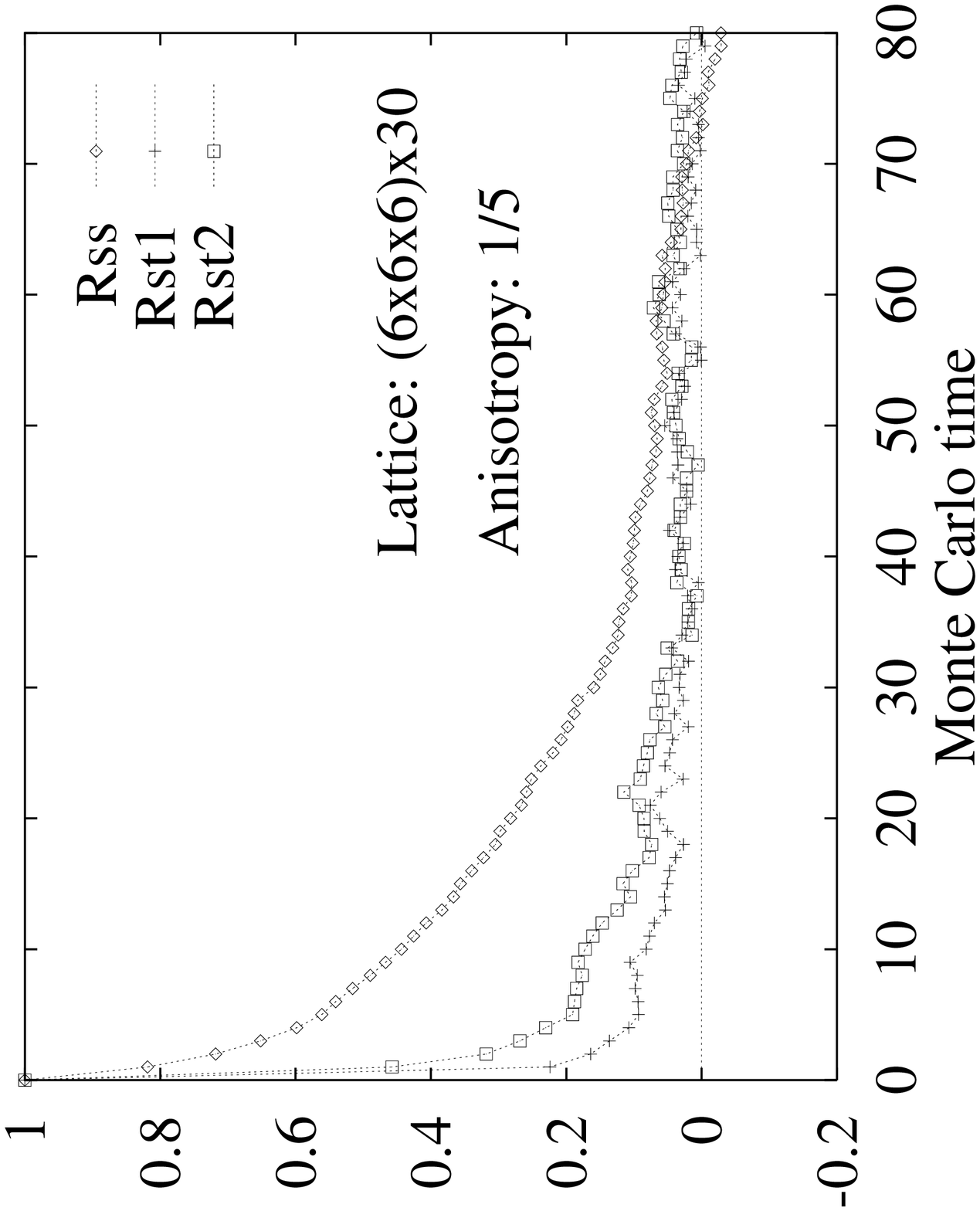}}
    \end{center}
  \end{minipage}
  \caption{Effects of anisotropy on the autocorrelation functions for the 
           $2 \times 1$ rectangles.}
  \label{arec}
\end{figure}

In Fig.~\ref{awil} and Fig.~\ref{arec} we present comparison graphs of the
autocorrelation functions for the simple plaquettes and the $2 \times 1$
rectangles. At small anisotropy of the lattice the autocorrelation times of the
spatial and temporal gluon operators are small and almost indistinguishable.


With increasing (inverse) anisotropy, $\rho^{-1}$, the degeneracy lifts. The
autocorrelation times increase for Pss and Rss, and remain small and degenerate
for Pst, Rst1 and Rst2 observables. This is to be expected, since Pss and Rss
couple strongly to $\xi^{\rm lat}_{t}$.

As can be observed from the graphs our autocorrelation functions distinguish
well between the spatial and temporal modes on the lattice, and can be a good
starting point for the optimization procedure.

\section{AUTOCORRELATION TIMES}

The integrated autocorrelation times for an observable $X$ are obtained using
the method proposed by Sokal~\cite{sok,jeg}, namely
\begin{equation}
  \tau_{\rm int} = \frac{1}{2} \sum^{n-m}_{t=-n+m}
                   \frac{R(t)}{R(0)}
\end{equation}
with
\begin{equation}
  R(t) = \frac{1}{n - \mid t \mid} \sum^{n - \mid t \mid}_{i = 1}
         (X_{i} - \bar{X})(X_{i + \mid t \mid} - \bar{X})
\end{equation}
and $m$ chosen so that $\tau_{\rm int} \ll m \ll n$. The smallest value of $m$
for which $m/\tau_{\rm int} \geq 4$ has been chosen in a self-consistent
manner. From here we derive an estimate for the error of $\tau_{\rm int}$ given
by the formula
\begin{equation}                 
  \sigma^{2}_{\tau_{\rm int}} = \frac{2(2m+1)}{n}\tau^{2}_{\rm int}
\end{equation}
We call this criterion -- ``criterion $m/4$''.

We also double check our results against the procedure employed by the QCD-TARO
Collaboration~\cite{for}, wherein the autocorrelation time is defined as:
\begin{equation}
  \tau_{\rm int} = \rho(0) + 2 \sum_{t=1}^{N} \rho(t) \frac{N-t}{N}
\end{equation}
where $N$ is determined so that $\tau_{\rm int}$ is maximized, but $N < 10 \%$
of the total sample and $N < 3 \tau_{\rm int}$.  Likewise, we call this
criterion -- ``criterion $\tau_{\rm max}$''.

We use hybrid overrelaxation updating scheme, which is considered to be the
state-of-the-art algorithm for bosonic systems when no cluster algorithm is
available~\cite{wolff}. It simply consists in the mixing of Cabibbo-Marinari
pseudo-heatbath and the Brown-Woch microcanonical overrelaxation sweeps with a
ratio $1 : n_{o}$~\cite{hb,over}.

\section{ANISOTROPY-DEPENDENCE OF $\!\tau_{\rm int}$}

\begin{figure}[htb]
  \begin{minipage}[t]{\hsize}
    \begin{center}
      \leavevmode
      \epsfysize=\hsize 
      \rotate[r]{\epsfbox{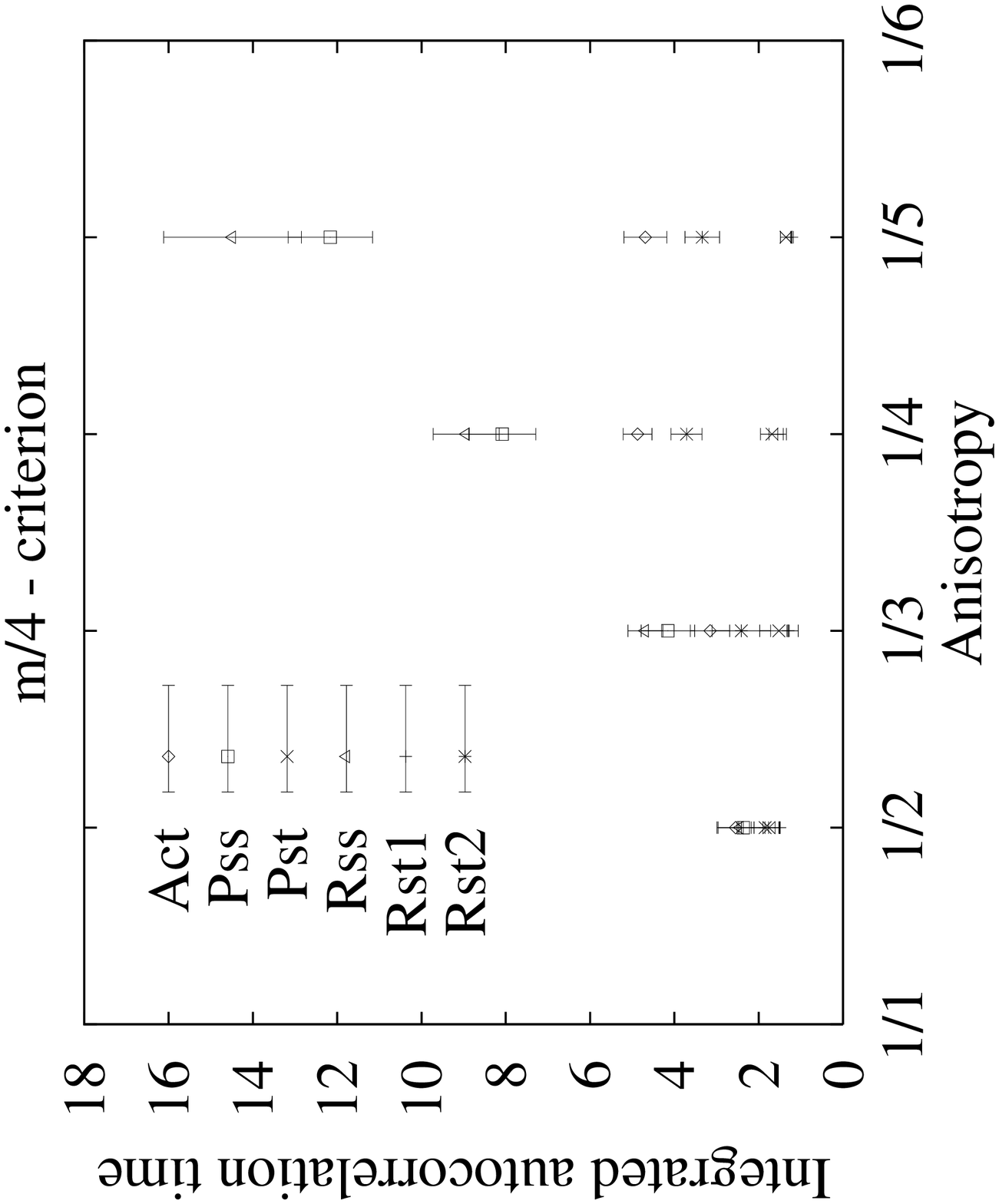}}
    \end{center}
  \end{minipage}
  \begin{minipage}[t]{\hsize}
    \begin{center}
      \leavevmode
      \epsfysize=\hsize 
      \rotate[r]{\epsfbox{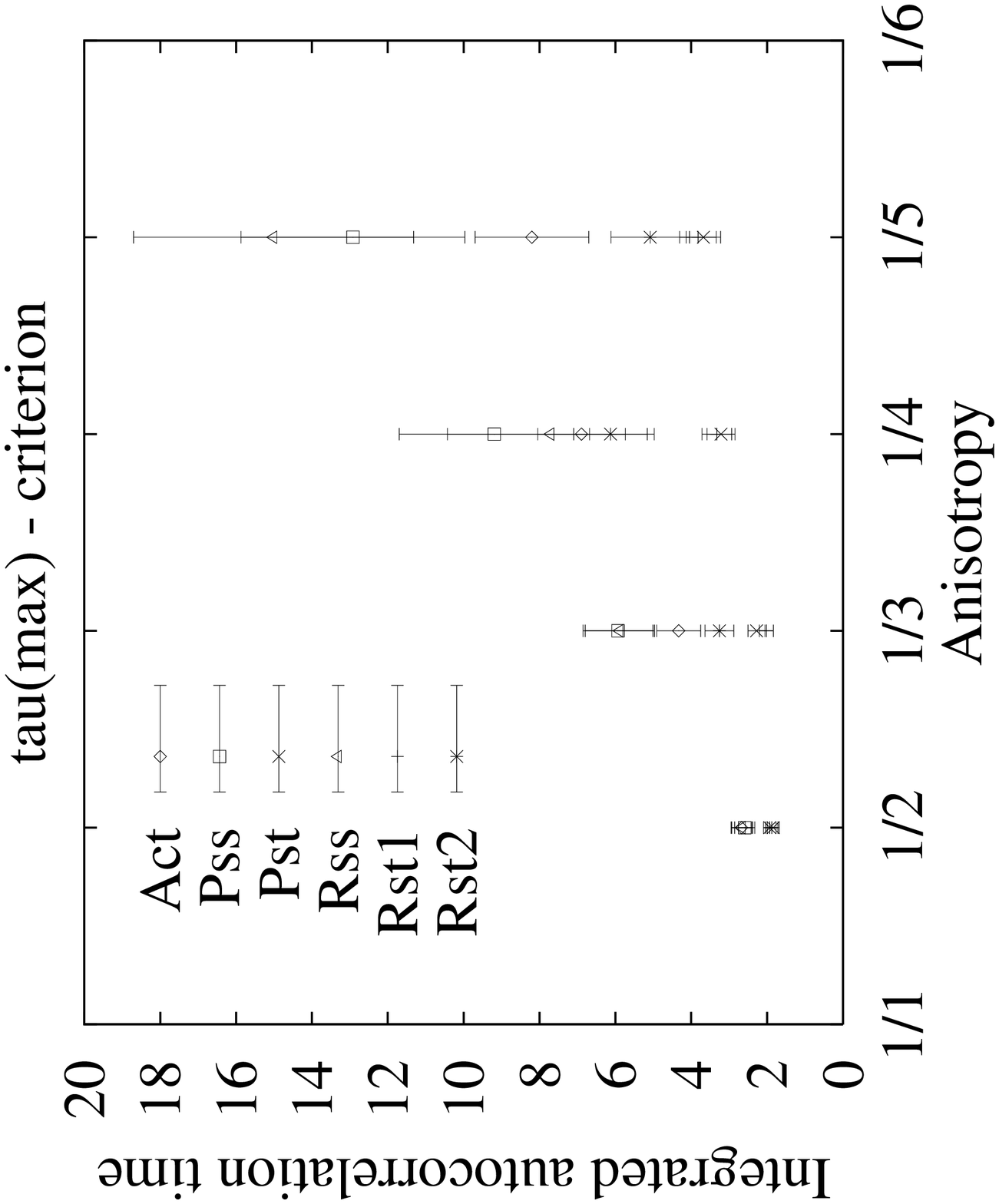}}
    \end{center}
  \end{minipage}
  \caption{Scaling behavior of $\tau_{\rm int}$ on anisotropy.}
  \label{err}
\end{figure}

In Fig.~\ref{err} we show our final results for the dependence of the
integrated autocorrelation time on the anisotropy of the lattice for the six
lattice observables. We find small variation of the amplitude of the integrated
autocorrelation time $\tau_{\rm int}$ for the temporal loops. The error bars
for these observables are small.  It is apparent that the modes are strongly
suppressed because they effectively propagate in the spatial domain with
characteristic correlation length $\xi^{\rm lat}_{s}$, which is small.  In
contrast, the integrated autocorrelation times for the spatial loops increase
as $\rho^{-1}$ increases.

%

\section{CONCLUSIONS}

This study is an initial attempt to perform accurate studies of QCD gluon
update dynamics on lattices with large spatial lattice spacing and small
temporal lattice spacing.

This work is at an early stage of development. We have found clear evidence
that the spatial and temporal link variables have different autocorrelation
behavior, and that this difference becomes more extreme as the anisotropy is
increased. We hope to look for an optimal update scheme for anisotropic
simulations.  This will be accomplished by exploiting the freedom we have on
such lattices to vary pseudo-heatbath and overrelaxation updates on the spatial
and temporal links independently. Since the action has explicitly broken
Euclidean symmetry, we can expect that an optimal update scheme will involve
different update methods for spatial and temporal links.  We anticipate the
optimal update method will involve overrelaxation of the spatial links (which
have been shown here to have the longest autocorrelation times) with heatbath
updates of temporal links to ensure the overall update is ergodic.

\end{document}